# Spin Hall Angle Quantification from Spin Pumping and Microwave Photoresistance

Z. Feng[1], J. Hu[1], L. Sun[1], B. You[1,*], D. Wu[1], J. Du[1], W. Zhang[1], A. Hu[1], Y. Yang[2], D. M. Tang[2], B. S. Zhang[2], H. F. Ding[1,*]

[1]National Laboratory of Solid State Microstructures and Department of Physics,

[2]School of Electronic Science and Engineering

Nanjing University, 22 Hankou Road, Nanjing 210093, P. R. China

*Email: youbiao@nju.edu.cn, hfding@nju.edu.cn

## Abstract

We present a method to quantify the spin Hall angle (SHA) with spin pumping and microwave photoresistance measurements. With this method, we separate the inverse spin Hall effect (ISHE) from other unwanted effects for permalloy/Pt bilayers using out-of-plane microwave excitation. Through microwave photoresistance measurements, the in- and out-of-plane precessing angles of the magnetization are determined and enabled for the exact determination of the injected pure spin current. This method is demonstrated with an almost perfect Lorentz line-shape for the obtained ISHE signal and the frequency independent SHA value as predicted by theory. By varying the Pt thickness, the SHA and spin-diffusion length of Pt is quantified as $0.012 \pm 0.001$ and $8.3 \pm 0.9\,\text{nm}$, respectively.

# Introduction

Electrons have two fundamental properties, the charge and the spin. Over the past, information technology has made tremendous progress even though mainly the charge property of the electrons was only used. One can image that the adding usage of the spin property will enrich the functionalities of the devices. More importantly, pure spin devices may provide a potential solution for the power consumption problem which becomes increasingly serious with the speed acceleration and size reduction of the micro-electronic devices.[1] The detection of a spin current however is not easy. Optical detection has been successfully used for observation of spin accumulation,[2-4] but this method, is limited to semiconductor systems which typically have long spin diffusion lengths. A more general solution is to convert the spin current to the charge current, which is the basis for existing technology. Therefore, the conversion of the spin and charge currents is one of the key issues for spintronics technology.

The spin Hall effect (SHE) refers to the generation of a spin current transverse to an applied charge current in a paramagnetic metal or a doped semiconductor.[5, 6] Concurrently, a spin current can also give rise to a transverse charge current, which is called the inverse spin Hall effect (ISHE). The efficiency of the spin-charge conversion can be quantified by a single material-specific parameter, *i.e.*, the spin Hall angle (SHA), $\theta_{SH}$. It is defined as the ratio of the spin Hall and charge conductivities.[7] The SHA can be measured through the nonlocal magneto-transport measurements[8-12] or the method based on spin pumping due to ferromagnetic resonance (FMR).[13-18] Because of the complexity of the interface effect, it is typically difficult to estimate the exact amplitude of the injected pure spin current with the first method. The second method is of more advantage as the above difficulty can be removed with additional FMR measurements. Surprisingly, the experimentally reported values are quite different for nominally identical materials, even for similar methods utilizing spin pumping. For instance, the measured SHA value for Pt varied between 0.0067 and 0.08.[15-19] With a literature value of the spin diffusion length, $\lambda_{sd}$ =10 nm, Mosendz et al., reported the SHA for Pt to be 0.0067 [15] and later refined it to 0.013 after correcting for the elliptical magnetization precessing trajectory.[16] Using the same spin diffusion length, the measurements of Ando et al., however, show a value of 0.04.[18] With the Pt thickness dependent measurements, Azevedo et al., obtained a SHA value of 0.04 (in their original paper, the value is 0.08. But their definition is a factor 2 larger than the one that commonly used) and a spin diffusion length of 3.7±0.2 nm.[17] The discrepancy may be related to the fact that the ISHE signal is typically mixed with the unwanted effects related to the anisotropic magnetoresistance (AMR) effect.[15-17] Therefore, the correct separation of the ISHE signal from the other effects is crucial for the SHA estimation. In addition, the measured ISHE voltage is closely related to the SHA and the amplitude of the injected pure spin current as well as the spin diffusion length. In such case, the correct measurements of the amplitude of the injected pure spin current and the spin diffusion length are also very essential. This, however, is not easy. For example, the effective microwave magnetic-field $h_{rf}$ acting on the magnetic layer can be different even with the same microwave power input, as it also depends on the thicknesses of both the ferromagnetic and nonmagnetic layers.

In this paper, we present a method to separate the ISHE from other effects for permalloy (Py)/Pt bilayers with an out-of-plane microwave excitation. The successful separation is demonstrated with an almost perfect Lorentz line-shape for the obtained signal and the frequency independent SHA value. Instead of using the microwave magnetic-field $h_{rf}$ to calculate the in- and out-of-plane precessing angles of the magnetization, we directly measure them through the microwave photoresistance measurement.[20, 21] This allows for the exact estimation of the injected pure spin current for individual samples. With varying the Pt thickness, the SHA and spin-diffusion length of Pt are quantified.

## Theory

The basic theory of utilizing the spin pumping effect for measuring the SHA has been described in Ref. 15, 16. For the reader's convenience, we briefly summarized it below. Spin pumping during the excitation of FMR occurs when the precession of the magnetization in a ferromagnetic (FM) layer injects a pure spin current into an adjacent nonmagnetic (NM) layer as shown in the schematic picture in Fig. 1(a).[22-24] Due to the inverse spin Hall effect, the injected pure spin current creates a transverse voltage, *i.e.*, the inverse spin Hall voltage induced by spin pumping $V_{ISHE}^{SP}$.[13-18] Taking into account the spin relaxation and diffusion in the NM layer, the *dc* part of pure spin current density along the y-direction can be written as:

$$j_s(y) = j_s^0 \frac{\sinh[(t_N - y)/\lambda_{sd}]}{\sinh(t_N/\lambda_{sd})} \qquad (1)$$

where $t_N$ and $\lambda_{sd}$ are the thickness and the spin-diffusion length of NM layer, respectively. $j_s^0$ is the spin-current density at the FM/NM interface (y = 0), and it is related to the effective spin mixing conductance $g_{eff}^{\uparrow\downarrow}$, the microwave frequency $f$ and the precessing angle of the FM. Following the basic theory of FMR, its magnetic field ($H$) dependence can be written as:

$$j_s^0(H) = \frac{\hbar}{2} g_{eff}^{\uparrow\downarrow} f \alpha_1 \beta_1 \frac{\Delta H^2}{(H - H_0)^2 + \Delta H^2} \qquad (2)$$

where $H_0$ is the resonance magnetic field, $\Delta H$ is the half-width of the FMR linewidth, $\alpha_1$ and $\beta_1$ are the maximum amplitudes of the in- and out-of-plane precessing angles of the magnetization, respectively. Due to the ISHE, the pure spin current $j_s(y)$ gives rise to a transverse charge current $\vec{j}_c(y) = \theta_{SH}\left(\dfrac{2e}{\hbar}\right) j_s(y)[\vec{n}\times\vec{\sigma}]$, where $\vec{n}$ is the direction of the pure spin current, $\vec{\sigma}$ is the polarization vector of the *dc* spin current. The charge current flowing along the NM layer (with length $L$, width $w$ and resistance $R_N$) generates a voltage, *i.e.*, $V_{ISHE}^{SP}$ along the z-direction. By integrating along the y-direction, the field dependence can be written as:

$$V_{ISHE}^{SP}(H) = R_N \int (\vec{j}_c(y)\bullet \hat{z})ds = \theta_{SH} \lambda_{sd} g_{eff}^{\uparrow\downarrow} fewR_N \tanh\left(\dfrac{t_N}{2\lambda_{sd}}\right)\alpha_1 \beta_1 \sin\alpha_0 \dfrac{\Delta H^2}{(H-H_0)^2 + \Delta H^2} \quad (3)$$

where $\alpha_0$ is the angle between $H$ and the z-axis, as shown in Fig. 1(a). From the above equation, we see that $V_{ISHE}^{SP}$ has a Lorentz line-shape signal as a function of $H$.

In any real measurements, $V_{ISHE}^{SP}$ is often accompanied by another voltage ($V_{AMR}$) due to anisotropic magnetoresistance. $V_{AMR}$ is the spin rectification voltage caused by the induction current $I_1 \cos(\omega t)$ and the oscillating resistance $R(t) = R_0 - R_A \sin^2[\alpha_0 + \alpha_1(\omega t)]$ caused by the AMR effect in the FM stripe.[15-17, 20] For out-of plane $h_{rf}$ (corresponding to our experimental setup), the voltage can be written as[20]:

$$V_{AMR}(H) = -\dfrac{I_1 R_A \alpha_1 \sin 2\alpha_0}{2}\left(\dfrac{\Delta H^2}{(H-H_0)^2 + \Delta H^2}\cos\phi - \dfrac{(H-H_0)\Delta H}{(H-H_0)^2 + \Delta H^2}\sin\phi\right) \quad (4)$$

where $R_A$ is the resistance difference when magnetization is parallel and perpendicular to the stripe, and $\phi$ is the phase difference between the *rf* current and the magnetization at resonance.

From the above discussion, we can find that the measured voltage signal can have two components, $V = V_{ISHE}^{SP} + V_{AMR}$. $V_{ISHE}^{SP}$ has a Lorentz line-shape, *i.e.,* it is symmetrical with respect to resonance field $H_0$, while $V_{AMR}$ contains both symmetric and asymmetric contributions. These characters make it difficult to separate both effects from the symmetry point of view. To quantify $\theta_{SH}$, we however need to distinguish $V_{ISHE}^{SP}$ from $V_{AMR}$ firstly. In addition, one needs to correctly obtain other unknown parameters described in Eq. (3), such as $g_{eff}^{\uparrow\downarrow}$、$\alpha_1$、$\beta_1$、$\lambda_{sd}$ etc.

With careful analysis, one can find that $V_{ISHE}^{SP}$ and $V_{AMR}$ have a different dependency with respect to $\alpha_0$. More specifically, $V_{ISHE}^{SP}$ is proportional to $\sin\alpha_0$ and $V_{AMR}$ is linearly dependent on $\sin 2\alpha_0$. Therefore, one can eliminate the $V_{AMR}$ contribution by choosing two specific geometries, $\alpha_0 = 90^0$ and $\alpha_0 = 270^0$, where $V_{AMR} = 0$, and $V_{ISHE}^{SP}$ reaches its maximum amplitude. In our measurements discussed below, we use this method to separate $V_{ISHE}^{SP}$ from the unwanted signals.

The effective spin mixing conductance $g_{eff}^{\uparrow\downarrow}$ can be determined by the enhanced Gilbert damping factor due to the losing spin momentum during spin pumping[13-18]:

$$g_{eff}^{\uparrow\downarrow} = \frac{4\pi M_0 t_F}{g\mu_B}(\alpha_{F/N} - \alpha_F) \qquad (5)$$

The damping factor of FM/NM layer $\alpha_{F/N}$, and FM layer $\alpha_F$ can be calculated from the half width of the FMR linewidth $\Delta H$ through: $\alpha = \Delta H \frac{\gamma}{2\pi f}$, where $\gamma$ is the gyromagnetic ratio.

The maximum amplitudes of in- and out-of-plane angles, $\alpha_1$ and $\beta_1$ can be determined by microwave photoresistance effect.[20, 21] Microwave photoresistance is the *dc* resistance change in the FMR. The magnetization precession alters the angle of the magnetization with respect to *dc* current, resulting a change of the time-averaged AMR. For a single FM layer, the field dependent microwave photoresistance is given as:[20]

$$\Delta R_{MW}^F(H) = -\frac{R_A}{2}\left(\alpha_1^2 \cos 2\alpha_0 + \beta_1^2 \cos^2\alpha_0\right)\frac{\Delta H^2}{(H-H_0)^2 + \Delta H^2} \qquad (6)$$

When $\alpha_0 = 90^0$ or $\alpha_0 = 270^0$, it can be simplified as:

$$\Delta R_{MW}^F(H) = \frac{R_A}{2}\alpha_1^2 \frac{\Delta H^2}{(H-H_0)^2 + \Delta H^2} \qquad (7)$$

Therefore, $\alpha_1$ can be calculated with equation (7) after we take measurements of $\Delta R_{MW}^F$. Besides, $\beta_1$ can also be determined according to FMR theory,

$$\frac{\alpha_1}{\beta_1} = \sqrt{1 + \frac{M_{eff}}{H_0}} \tag{8}$$

where $M_{eff}$ is the effective magnetization of FM. The real sample is a double layer which consists of both FM and NM layers. For this, one needs to make a correction with the assumption of parallel resistance configuration. At last, the spin diffusion length $\lambda_{sd}$, together with the spin Hall angle $\theta_{SH}$, need to be obtained with the NM thickness dependent measurements and the fitting according to Eq. (3) as we will discuss below.

**Experiments**

As shown in Fig. 1 (b), the Py/Pt bilayer stripes (in light gray color) are integrated into the slots between the signal and ground lines (in brown color) of a coplanar waveguide (CPW). In this configuration, the magnetic dynamics are excited with an out-of-plane microwave magnetic field $h_{rf}$. We note that our experimental configuration is similar to the spin dynamo described by Gui et al.[25] The stripes' lateral dimensions are 2.5 mm*20 μm, and the thickness of Py layer is fixed at 16 nm while the thickness of Pt layer varies from 2 nm to 65 nm. The bilayer stripes are prepared by photolithography, magnetron sputtering deposition, and lift-off on semi-insulating GaAs substrates. The Pt and Py thicknesses are calibrated with X-ray diffraction. Subsequently, a copper CPW with a 50 Ω characteristic impendence and the electrical contacts are fabricated. The measurements are performed at room temperature (RT). In order to achieve high sensitivity, a lock-in technique is used. A vector network analyzer (VNA) supplies to the CPW a CW microwave, which is modulated with a 51.73 kHz signal. The lock-in amplifier picks up the voltage signal as a function of external magnetic field $H$. The magnetic field $H$ with controllable field strength can be rotated within the film plane. For the microwave photoresistance measurements, a constant *dc* current is applied to the stripe through Keithley 2400 SourceMeter. A 50 kΩ resistor is used in series with the SourceMeter to minimize the flowing of the *ac* voltage signal into the source branch.

## Results and discussions

Before showing the experimental results, we will first discuss the criteria for the pure ISHE measurements. As discussed in the section of theory, the measured voltage typically contains two parts, $V_{ISHE}^{SP}$ and $V_{AMR}$. These two voltages bear different characters due to their different physical origins. From Eq. (3) and Eq. (4), we learn that $V_{ISHE}^{SP}$ has a Lorentz line-shape and its magnetic field dependence should be symmetric with respects to the resonance field while $V_{AMR}$ contains both symmetric and asymmetric parts. Secondly, $V_{ISHE}^{SP}$ is proportional to $\sin\alpha_0$ and $V_{AMR}$ is linearly dependent on $\sin 2\alpha_0$. Therefore, we performed the measurements for $\alpha_0 = 90^0$ and $\alpha_0 = 270^0$, where $V_{ISHE}^{SP}$ reaches its maximum amplitude and $V_{AMR} = 0$. The measured signal for these two configurations should bear the same magnitude but with opposite sign with the same excitation. Thirdly, $V_{ISHE}^{SP}$ is generated by the *dc* component of the injected pure spin current while $V_{AMR}$ is proportional to the induction current and it is frequency dependent. Therefore, we set up 4 criteria to determine the pure ISHE signal: (i) the field dependence of measured voltage should have a Lorentz line-shape; (ii) it has opposite sign for $\alpha_0 = 90^0$ and $\alpha_0 = 270^0$; (iii) it has the same amplitude for $\alpha_0 = 90^0$ and $\alpha_0 = 270^0$ with the same injected pure spin current; (iv) and most importantly, as $\theta_{SH}$ is a material specific parameter, it should be independent on the microwave frequency used for the measurements. Particularly，the ISHE signal we measured is generated by the *dc* component of the injected pure spin current.

Figure 2(a) shows a typical result of the measured *dc* voltages *V* as a function of *H* for $\alpha_0 = 90^0$ and $\alpha_0 = 270^0$ with the same microwave power input. The symbols are the experimental data while the lines are the Lorentz fit. The sample is Py(16nm)/Pt(15nm) and the frequency of the microwave is fixed to 8 GHz. We find that both curves are close to the Lorentz line-shape and they have opposite sign for $\alpha_0 = 90^0$ and $\alpha_0 = 270^0$. Therefore, the above discussed criterion (i) and (ii) are satisfied, suggesting the obtained signals are mainly caused by the ISHE. One however can also find that the curves have different amplitudes for $\alpha_0 = 90^0$ and $\alpha_0 = 270^0$. This is, at first glance, in contrast with the criterion (iii) even though the difference between them is not big. As discussed above, the criterion (iii) requires a precondition that the injected pure spin currents are the same for these two configurations. Even with the same microwave power input, this precondition may not be necessary fulfilled as will be discussed below. From Eq. (2), we can find that the injected pure spin current is proportional to the product of the effective spin mixing conductance, the microwave frequency and in- and out-of-plane precessing angles, *i.e.*, $g_{eff}^{\uparrow\downarrow} f \alpha_1 \beta_1$. For a given sample, the spin mixing conductance and the microwave frequency are fixed, the precessing angles, $\alpha_1$ and $\beta_1$ can be determined by the microwave photoresistance measurements,[20, 21] as also discussed in Eq. (6)-(8).

The microwave photoresistance measurements are performed for the same Py/Pt sample at $\alpha_0 = 90^0$ and $\alpha_0 = 270^0$ with the same input microwave power used for the above measurements. To eliminate the influence of $V_{ISHE}^{SP}$, additional positive/negative *dc* currents $I_0$ (=2.5mA) are used, and microwave photoresistance of Py/Pt bilayers is obtained by their difference: $\Delta R_{MW}^{F/N} = [V(I_0) - V(-I_0)]/2I_0$. In addition, the modulation frequency 51.73 kHz is high enough to exclude any bolometric effect.[21] Figure 2 (b) shows the results of $\Delta R_{MW}^{F/N}$ as a function of the applied magnetic field $H$. The symbols are the experimental data while the lines are the Lorentz fit. They also have a Lorentz line-shape, corresponding to Eq. (7), and more importantly, they are different for $\alpha_0 = 90^0$ and $\alpha_0 = 270^0$, suggesting the precessing angles for these two configurations are different even though the same microwave power is used. We can find that $\Delta R_{MW}^{F/N}\big|_{270^0} > \Delta R_{MW}^{F/N}\big|_{90^0}$, which is consistent with the relationship for the magnitude of the measured voltages for these two configurations. To be more quantitative, we need to calculate $\alpha_1$ and $\beta_1$. For this, one need to obtain the microwave photoresistance of the single FM stripe, $\Delta R_{MW}^{F}$. This can be calculated from $\Delta R_{MW}^{F/N}$ through the shunt relationship: $\Delta R_{MW}^{F} \approx \frac{\Delta R_{MW}^{F/N}(R_F+R_N)^2}{R_N^2}$ by assuming a parallel connection (Through comparing the angular dependent resistance measurements for both pure Py and Py/Pt film, we find this assumption only gives an uncertainty of 5% for samples with the Pt thickness above 2 nm). $R_F$ is the resistance of FM layer when its magnetization is perpendicular to the stripe, and $R_N$ can be calculated from $R_F$ and $R_{F/N}$ (resistance of FM/NM bilayer) through the shunt relationship. $R_F$、$R_{F/N}$ and $R_A$ are obtained by four probe static magnetoresistance measurements. For this particular sample, we obtain that: $R_A = 36.3\Omega$, $R_F = 2700\Omega$, and $R_{F/N} = 1179\Omega$. With the shunt relationship discussed above, we can calculate $R_N = 2093\Omega$. From the Lorentz fit of the two curves in Fig. 2(b), we obtain $\Delta R_{MW}^{F/N}\big|_{270^0} = 0.366 m\Omega$, $\Delta R_{MW}^{F/N}\big|_{90^0} = 0.286 m\Omega$. With the resonance field $H_0 = 705 Oe$ and the effective magnetization $\mu_0 M_{eff} = 0.967 T$, we can further calculate the precessing angle according to Eq. (6)-(8). We find that $\alpha_1 = 0.52^0$, $\beta_1 = 0.14^0$ for $\alpha_0 = 90^0$ and $\alpha_1 = 0.59^0$, $\beta_1 = 0.15^0$ for $\alpha_0 = 270^0$. Therefore, we further normalized the measured voltage to the precessing angle. Interestingly, we find that $\frac{V}{\alpha_1 \beta_1}\big|_{90^0} \approx -\frac{V}{\alpha_1 \beta_1}\big|_{270^0}$ and the criterion (iii) is fulfilled.

Following the method given in Ref. [20], we can also estimate the effective microwave magnetic-field $h_{rf}$ acting on FM layer from $\alpha_1 = \frac{h_{rf}}{\alpha(2H_0 + M_{eff})}$. With the Gilbert damping factor $\alpha = 0.011$, we obtain $h_{rf} \approx 1.11 Oe$ for $\alpha_0 = 90^0$ and $h_{rf} \approx 1.26 Oe$ for $\alpha_0 = 270^0$. The exact reason for the different $h_{rf}$ for these two configurations is unclear at present stage. We find this difference also exists for the single Py film, and it is independent on $I_0$ that used for the resistance measurements, suggesting it is not caused by the current induced heating. We note that similar effects have also been shown in other systems.[26, 27]

To eliminate the residual AMR effects caused by the small experimental misalignment and the contact rectification effect,[28] we redefine a normalized ISHE caused by spin pumping:

$$\tilde{V}_{ISHE}^{SP} = \left( \left.\frac{V}{\alpha_1 \beta_1}\right|_{90^0} - \left.\frac{V}{\alpha_1 \beta_1}\right|_{270^0} \right) / 2 \tag{9}$$

When $H$ is not perfectly applied perpendicular to the strip, $V_{AMR}$ signal will be mixed in the measurements. With $\tilde{V}_{ISHE}^{SP}$, the mixing effect can be minimized because of $V_{AMR}(\alpha_0) = V_{AMR}(\alpha_0 + 180^0)$. In addition, there exists a small voltage due to the contact rectification effect.[28] This additional small voltage bears similar symmetry as $V_{AMR}$ and it can also be eliminated by $\tilde{V}_{ISHE}^{SP}$ as well. A typical field dependence of $\tilde{V}_{ISHE}^{SP}$ is presented in Fig. 3(a). The solid symbols are the experimental data and the line is the Lorentz fit. We can find that it shows an almost perfect Lorentz line-shape. This strongly supports its spin pumping origin.

Above, we discussed the criteria (i)-(iii). In the following, we will continue to discuss the criterion (iv), *i.e.*, the frequency independence of $\theta_{SH}$. For this, we further performed the measurements for the same sample but with different frequencies. As $\theta_{SH}$ is frequency independent, we can find, from Eq. (3), that $V_{ISHE}^{SP}$ has 3 frequency-dependent parameters for a given sample: $f$, $\alpha_1$ and $\beta_1$. The normalized ISHE caused by the spin pumping, however, only has a simple linear dependence with the frequency, *i.e.*, $\tilde{V}_{ISHE}^{SP}(H_0) = \theta_{SH} \lambda_{sd} \tanh\left(\frac{t_N}{2\lambda_{sd}}\right) g_{eff}^{\uparrow\downarrow} e w R_N f$. Figure 3(b) shows the dependence of the measured $\frac{\tilde{V}_{ISHE}^{SP}(H_0)}{eR_N w}$ as a function of the frequency for two samples: Py(16nm)/Pt(15nm) and Py(16nm)/Pt(6nm), respectively. We can find that both curves show almost perfect linear dependence, strongly supporting the frequency independence of $\theta_{SH}$. Therefore, the 4 criteria mentioned above are all fulfilled. The satisfaction of these 4 criteria also proves that our measurements for $V_{ISHE}^{SP}$, $\alpha_1$ and $\beta_1$ are correct. For the NM thickness dependent measurements discussed below, these 4 criteria are examined for all samples.

The effective spin mixing conductance $g_{eff}^{\uparrow\downarrow}$ can be determined by the enhanced Gilbert damping factor due to the losing spin momentum during spin pumping.[13-18] Figure 4(a) shows the $t_N$ dependence of $g_{eff}^{\uparrow\downarrow}$ calculated with Eq. (5). We can find that $g_{eff}^{\uparrow\downarrow}$ is saturated when $t_N \approx 2$ nm for Pt. This is consistent with the results of Ref. 29 where $g_{eff}^{\uparrow\downarrow}$ reaches saturation at ~1.5 nm for samples Cu/Py(3nm)/Cu(10nm)/Pt($t_N$)/Cu. The low saturation thickness shows that Pt is an effective spin sink material as pointed out by Tserkovnyak et al.[24] The obtained $g_{eff}^{\uparrow\downarrow} = 2.47(\pm 0.15) \times 10^{19} m^{-2}$ for the Pt thickness above 2 nm, is similar with the results from other groups.[15-18] We note that the saturation distance for $g_{eff}^{\uparrow\downarrow}$ is not the spin diffusion length $\lambda_{sd}$ as it describes the thickness of the NM layer which is necessary to sink the spin accumulation at the interface.[24, 29]

Up to now, the only parameters left unknown are $\theta_{SH}$ and $\lambda_{sd}$. They can be obtained through the NM thickness dependent measurements as we can easily derive that $\theta_{SH} \lambda_{sd} \tanh\left(\dfrac{t_N}{2\lambda_{sd}}\right) = \dfrac{\tilde{V}_{ISHE}^{SP}(H_0)}{eR_N w g_{eff}^{\uparrow\downarrow} f}$, where the right side of the equation can be measured experimentally. The results are shown in Fig. 4(b) for two frequencies. The solid circles and open squares are the experimental measured $\dfrac{\tilde{V}_{ISHE}^{SP}(H_0)}{eR_N w g_{eff}^{\uparrow\downarrow} f}$ for the frequency of 8 GHz and 9 GHz, respectively. We fit the data with $\theta_{SH}$ and $\lambda_{sd}$ as the only two parameters using the above mentioned relationship. The results are plotted as the solid (8 GHz) and the dash (9 GHz) line, respectively. The fitting yields $\theta_{SH} = 0.0120 \pm 0.0006$ and $\lambda_{sd} = 8.3 \pm 0.5$ nm for $f = 8$ GHz and $\theta_{SH} = 0.0118 \pm 0.0005$ and $\lambda_{sd} = 8.2 \pm 0.5$ nm for $f = 9$ GHz. We can find that their differences are within 2%, strongly supporting the frequency independence of the SHA. With the overall experimental error margin analysis, we obtain $\theta_{SH} = 0.012 \pm 0.001$ and $\lambda_{sd} = 8.3 \pm 0.9$ nm, respectively. The obtained spin diffusion constant is in good agreement with the non-local spin valve measurements[12] and the theoretical calculation[30]. We further calculate the spin Hall conductivity with the obtained SHA and the experimental determined the conductivity of Pt: $\sigma_{Pt} = (4.3 \pm 0.2) \times 10^6$ $(\Omega m)^{-1}$ according to $\sigma_{Pt}^{SH} = \theta_{SH} \bullet \sigma_{Pt}$. The value is $516 \pm 30$ $(\Omega cm)^{-1}$ and it is larger than the theoretical value of 330 $(\Omega cm)^{-1}$,[31] suggesting the existence of the extrinsic effect in this particular system. This can be understood as the calculation is made for single crystalline sample while our samples are polycrystalline films.

In the following, we will make a brief comparison of our results with those from other groups.[15-18] Ref. 15-16 successfully demonstrated the applicability of using spin pumping for spin Hall angle measurements. They disentangled $V_{ISHE}^{SP}$ and $V_{AMR}$ by assuming $V_{AMR}$ has only asymmetric component in their particular geometry. The SHA value obtained by them is very close to our result, suggesting the assumption may be valid. The measured magnetic field dependent ISHE voltage in Ref. 18 shows a Lorentz line-shape by placing the sample near the center of a TE$_{011}$ cavity. The authors calculated $\theta_{SH}$ to be 0.04 from a single NM thickness measurement and utilizing a literature value of $\lambda_{sd} = 10$ nm. Ref. 17 separated $V_{ISHE}^{SP}$ and $V_{AMR}$ with the angular dependent analysis and performed the NM thickness dependent measurements. The authors, however, did not measure the NM thickness dependent $g_{eff}^{\uparrow\downarrow}$. Instead, they fitted the data for the additional damping parameter with the formula mainly valid for diffusive material (Eq. (6) in the original paper). Pt, however, is an effective spin sink material[24] and $g_{eff}^{\uparrow\downarrow}$ is already saturated at about 2 nm as discussed above.

Moreover, Ref. 17 and Ref. 18 used the input microwave power to calculate the effective $h_{rf}$ and further estimated the precessing angles and the injected pure spin current. Our measurements show $h_{rf}$ can be different even the same microwave power is used. Therefore, we measured the precessing angles directly utilizing the microwave photoresistance effect for all the samples, which should give better estimation for the injected pure spin current.

## Summary

In summary, we find that the ISHE induced by spin pumping is typically mixed with signals due to AMR and the injected pure spin current is geometry and sample thickness dependent even for the same microwave power input. We develop a method to separate the ISHE signal from other unwanted signals using out-of-plane microwave excitation and determine the in- and out-of-plane precessing angles through the microwave photoresistance measurements. This method enables the exact quantification of the SHA. It is demonstrated for the Py/Pt bilayer system with an almost perfect Lorentz line-shape for the obtained ISHE signal and a frequency independent SHA value as expected from theory. By varying the Pt thickness, the SHA and spin-diffusion length of Pt is quantified as $0.012 \pm 0.001$ and $8.3 \pm 0.9$ nm respectively.

## Acknowledgments

The authors acknowledge Q.Y. Zhao and T. Jia for wire-bonding help. This work is supported by the State Key Program for Basic Research of China (Grants No. 2010CB923401), NSFC (Grants Nos. 10834001, 10974087, 11023002 and 11174131) and PAPD.

# Figure captions

Fig. 1 (a) Schematic illustration of the inverse spin Hall effect induced by spin pumping in a FM/NM bilayer system. (b) Experimental setup for spin pumping induced inverse spin Hall effect and microwave photoresistance measurements.

Fig. 2 (a) The measured magnetic field dependent *dc* voltages $V$ ($f$ = 8GHz) for the sample of Py(16nm)/Pt(15nm) at $\alpha_0 = 90^0$ (solid square) and $\alpha_0 = 270^0$ (open circle). (b) The magnetic field dependent microwave photoresistance for the same sample at $\alpha_0 = 90^0$ (solid square) and $\alpha_0 = 270^0$ (open circle) under the same microwave power input.

Fig. 3 (a) Normalized spin pumping voltage corresponding to Fig. 2(a). Solid squares are the experimental data and the line is the Lorentz fit. (b) Microwave frequency dependent $\frac{\tilde{V}_{ISHE}^{SP}(H_0)}{eR_N w}$ for Py(16nm)/Pt(15nm) (solid square) and Py(16nm)/Pt(6nm) (open circle), respectively.

Fig.4 (a) Pt thickness dependent $g_{eff}^{\uparrow\downarrow}$ for Py(16nm)/Pt($t_N$). The line is the guide for eyes. The value is saturated at $t_N \approx 2nm$ (b) Experimental determined Pt thickness dependent $\frac{\tilde{V}_{ISHE}^{SP}(H_0)}{eR_N w g_{eff}^{\uparrow\downarrow} f}$ at $f$ =8 GHz (solid circles) and $f$ = 9 GHz (open square). The lines are the fitting curves according to the formula: $\frac{\tilde{V}_{ISHE}^{SP}(H_0)}{eR_N w g_{eff}^{\uparrow\downarrow} f} = \theta_{SH} \lambda_{sd} \tanh\left(\frac{t_N}{2\lambda_{sd}}\right)$.

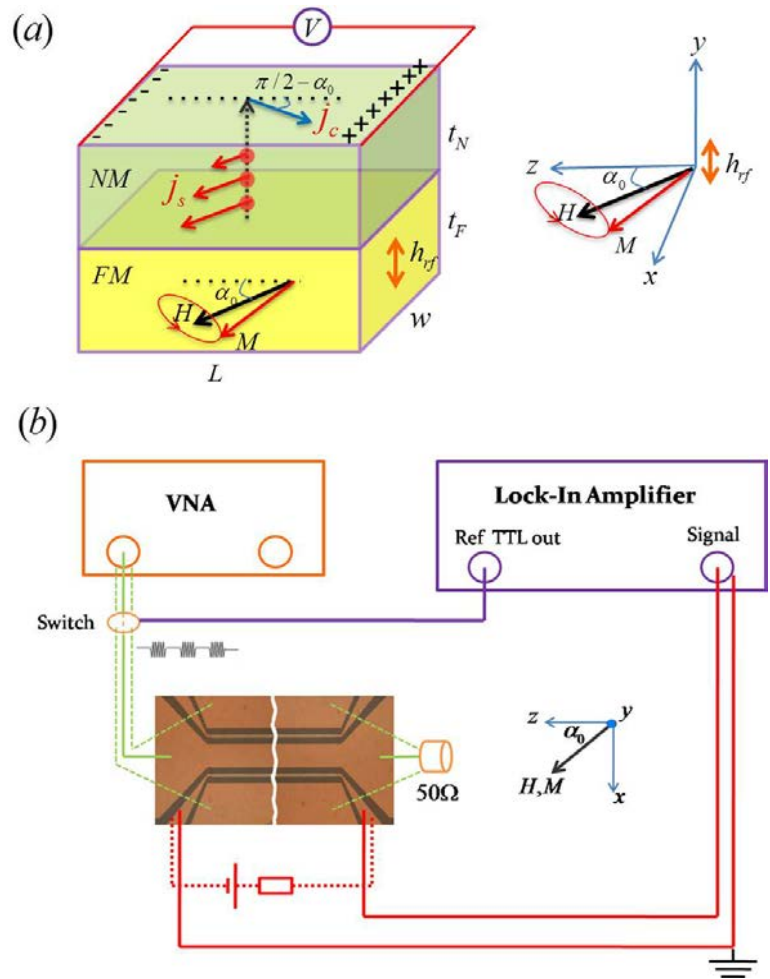

**Fig. 1 Feng et al.**

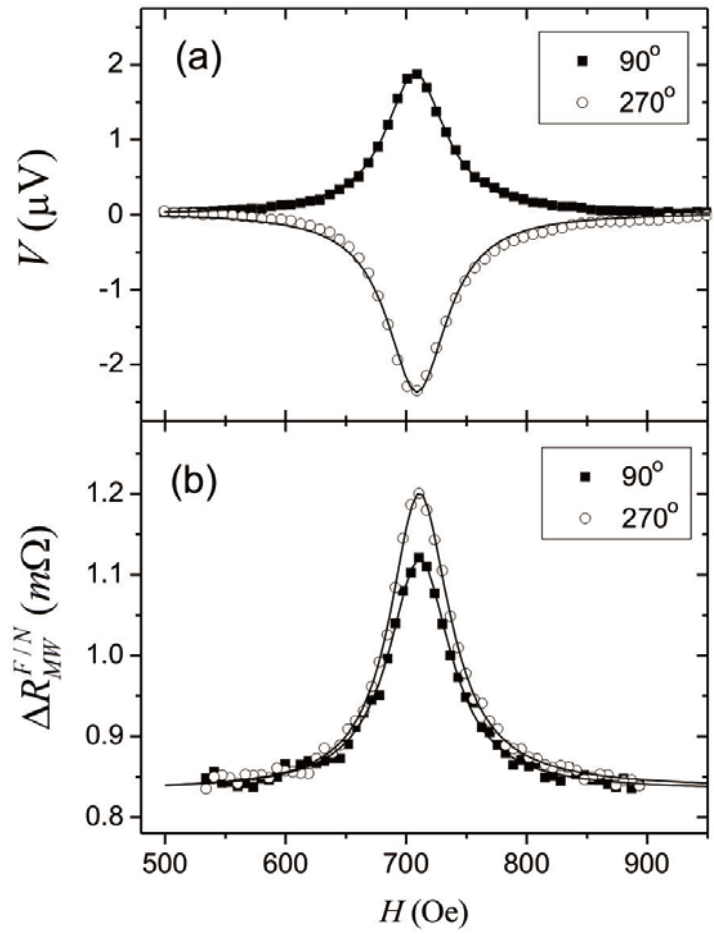

Fig. 2 Feng et al.,

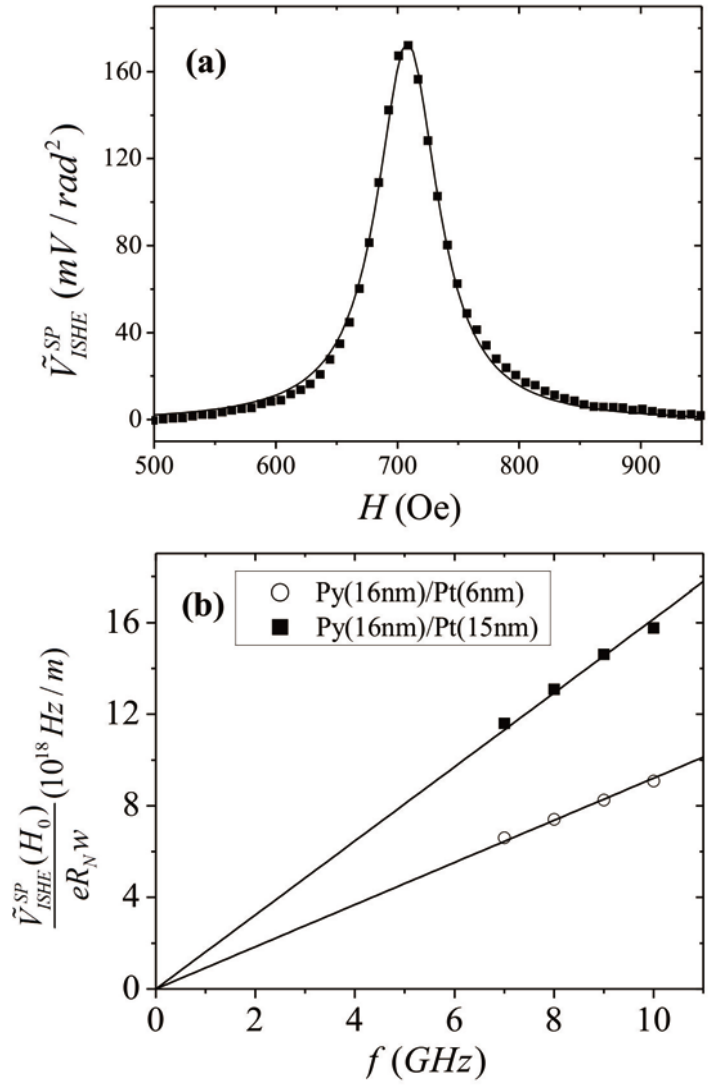

Fig. 3 Feng et al.,

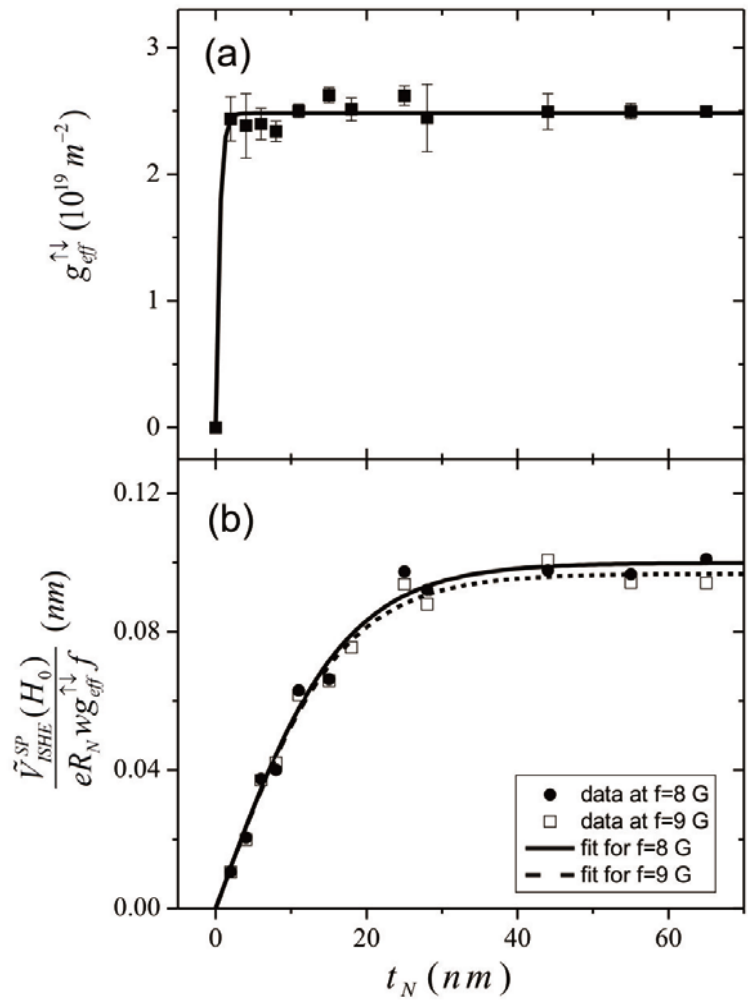

Fig. 4 Feng et al.,